\documentclass{aa}

\usepackage[varg]{txfonts}
\usepackage{epsfig,graphicx,natbib,url,twoopt}
\usepackage[varg]{txfonts}
\usepackage{hyperref}          
\hypersetup{
 colorlinks=true,  
 urlcolor=blue,    
 linkcolor=red,     
 citecolor=blue 
}

\setcitestyle{citesep={,}} 

\def\kms{\hbox{km$\;$s$^{-1}$}}

\def\Halpha{\mbox{H\hspace{0.1ex}$\alpha$}}
\def\Hbeta{\mbox{H\hspace{0.1ex}$\beta$}}

\def\CaK{\ion{Ca}{ii}~K}

\def\MgH{\ion{Mg}{ii}~h}

\def\IRIS{{IRIS}}

\begin{document}

\title{Signatures of ubiquitous magnetic reconnection in the deep atmosphere of sunspot penumbrae}

\author{Luc H. M. Rouppe van der Voort\inst{1,2} 
\and
Jayant Joshi\inst{1,2} 
\and
Vasco M. J. Henriques\inst{1,2} 
\and
Souvik Bose\inst{1,2} 
}
\authorrunning{L. Rouppe van der Voort et al.}

\institute{Institute of Theoretical Astrophysics,
  University of Oslo, %
  P.O. Box 1029 Blindern, N-0315 Oslo, Norway
\and
 Rosseland Centre for Solar Physics,
  University of Oslo, %
  P.O. Box 1029 Blindern, N-0315 Oslo, Norway
  }

\date{\today\ - submitted to A\&A December 18, 2020 / accepted January 26, 2021.}


\abstract
{
Ellerman bombs are regions with enhanced Balmer line wing emission and mark magnetic reconnection in the deep solar atmosphere in active regions and quiet Sun. 
They are often found in regions where opposite magnetic polarities are in close proximity. 
Recent high resolution observations suggest that Ellerman bombs are more prevalent than thought before.
}
{
We aim to determine the occurrence of Ellerman bombs in the penumbra of sunspots.
}
{
We analyze high spatial resolution observations of sunspots in the Balmer \Halpha\ and \Hbeta\ lines as well as auxiliary continuum channels obtained with the Swedish 1-m Solar Telescope and apply the $k$-means clustering technique to systematically detect and characterize Ellerman Bombs.
}
{
Features with all the defining characteristics of Ellerman bombs are found in large numbers over the entire penumbra.
The true prevalence of these events is only fully appreciated in the \Hbeta\ line due to highest spatial resolution and lower chromospheric opacity.
We find that the penumbra hosts some of the highest Ellerman bomb densities, only surpassed by the moat in the immediate surroundings of the sunspot. 
Some penumbral Ellerman bombs show flame morphology and rapid dynamical evolution.
Many penumbral Ellerman bombs are fast moving with typical speed of 3.7~\kms\ and sometimes more than 10~\kms.
Many penumbral Ellerman bombs migrate from the inner to the outer penumbra over hundreds of km and some continue moving beyond the outer penumbral boundary into the moat.
Many penumbral Ellerman bombs are found in the vicinity of regions with opposite magnetic polarity. 
} 
{
We conclude that reconnection is a near continuous process in the low atmosphere of the  penumbra of sunspots as manifest in the form of penumbral Ellerman bombs. These are so prevalent that they may be a major sink of sunspot magnetic energy.
}


\maketitle

\section{Introduction}
\label{sec:introduction}

Magnetic reconnection is a fundamental process in magnetized astrophysical plasmas for which magnetic energy is dissipated and converted into heat.
In the lower solar atmosphere, the hydrogen Balmer lines provide effective tracers of reconnection sites as they exhibit remarkable enhanced emission in their extended line wings as result of localised heating.
This phenomenon of enhanced wing emission, referred to as Ellerman ``bombs''
\citep[EBs,][]{1917ApJ....46..298E}, 
is particularly pronounced in emerging active regions with vigorous magnetic flux emergence.
At locations where opposite polarities are in close proximity (i.e., at the polarity inversion line), EBs appear as subarcsecond sized brightenings in \Halpha\ line wing 
\citep[see, e.g.,][]{2002ApJ...575..506G, 
2004ApJ...614.1099P, 
2006ApJ...643.1325F, 
2007A&A...473..279P, 
2008PASJ...60..577M, 
2008ApJ...684..736W} 
and \Hbeta{} line wing 
\citep{2017A&A...598A..33L, 
2020A&A...641L...5J} 
images.
The fact that the enhancement is only in the wings and that the EBs are invisible in the \Halpha\ line-core  locate  the  height of the reconnection  below  the  chromospheric canopy of fibrils 
\citep{2011ApJ...736...71W, 
2013ApJ...774...32V, 
2013ApJ...779..125N}. 
When observed from an inclined observing angle, sufficiently away from the center of the solar disk, and at sufficient spatial resolution, \Halpha\ wing images show EBs as tiny (1--2~Mm), bright, upright flames that flicker rapidly on a time scale of seconds \citep{2011ApJ...736...71W, 
2013JPhCS.440a2007R, 
2015ApJ...798...19N}. 
There is considerable spread in EB lifetimes but they rarely live longer than a few minutes.
We refer to 
\citet{2013JPhCS.440a2007R} 
and 
\citet{2019A&A...626A...4V} 
for recent reviews of observational EB properties and their visibility in different spectral diagnostics.

Traditionally, EBs have been associated with strong magnetic field environments and therefore regarded as a typical active region phenomenon. 
This view changed when 
\citet{2016A&A...592A.100R} 
and later 
\citet{2018MNRAS.479.3274S} 
reported the existence of tiny ($\lesssim0\farcs5$) Ellerman-like brightenings in quiet Sun when observed at extremely high spatial resolution.
\citet{2017ApJ...845...16N} 
found cases of quiet Sun EBs (QSEB) that were also visible in UV channels, suggesting that at least some QSEBs are energetic enough to become detectable in higher energy diagnostics. 
New high spatial resolution quiet Sun \Hbeta\ observations presented by 
\citet{2020A&A...641L...5J} 
show that QSEBs are much more ubiquitous than the lower spatial resolution \Halpha\ observations suggested. 
The shorter wavelength \Hbeta\ line allows for higher spatial resolution and higher temperature sensitivity and the observations suggest that about half a million QSEBs are present in the solar atmosphere at any time. 

The interpretation of EBs as markers of small-scale magnetic reconnection in the lower solar atmosphere has been reinforced by the advanced numerical simulations of 
\citet{2017ApJ...839...22H, 
2019A&A...626A..33H} 
and \citet{2017A&A...601A.122D}. 
In these simulations, heating occurs along current sheets that extend over several scale heights from the photosphere into the chromosphere. 
In synthetic \Halpha\ wing images, these current sheets are at the core of flame like structures that resemble the characteristic EB flames in observations. 

The sunspot penumbra is another environment in the lower solar atmosphere where magnetic reconnection is likely to occur.
In the penumbra, harboring an ``uncombed'' magnetic field topology with strong magnetic fields at highly variable inclination angles
and considerable dynamic forcing from convective flows, one may arguably expect ample occurrences of magnetic fields with differing angles at sufficient close proximity to effectively interact and reconnect
\citep[for reviews on the sunspot magnetic structure with strong-field vertical spines and and weaker-field horizontal inter-spines, see e.g., ][]{2011LRSP....8....4B, 
2017arXiv171207174T}. 
%
\citet{2013A&A...553A..63S} 
detected small regions of opposite polarity in a sunspot penumbra
\citep[also see][]{2013A&A...549L...4R, 2013A&A...550A..97F}, 
and that these regions harbor convective downflows. 
\citet{2015A&A...583A.119T} 
found ample regions with polarity opposite to the dominant sunspot polarity in a high quality Hinode SOT/SP map.

Based on the experience that EBs are often found at the interface between photospheric opposite polarity patches, we searched for EB signatures in high quality \Halpha\ and \Hbeta\ sunspot observations. 
In particular, we concentrated on the presence of flames in limbward observations as the telltale EB signature.
After close inspection of 13~datasets acquired over more than a decade of observation campaigns, we conclude that EBs are prevalent in sunspot penumbrae. 
The signature of penumbral EBs (PEBs) however is often subtle and requires excellent observing quality.
The \Hbeta\ line offers more clear detection in comparison to \Halpha, where the EB spectral signature is often hidden by dense superpenumbral filaments.
In this paper, we present results from analysis of the best datasets.

\section{Observations}
\label{sec:observations}

The observations were obtained with the Swedish 1-m Solar Telescope 
\citep[SST, ][]{2003SPIE.4853..341S} 
on the island of La Palma, Spain.
We used the 
CRisp Imaging SpectroPolarimeter 
\citep[CRISP,][]{2008ApJ...689L..69S} 
and the CHROMospheric Imaging Spectrometer (CHROMIS)
to perform imaging spectrometry
in the \Halpha\ and \Hbeta\ spectral lines.
We used the standard SST data reduction pipelines 
\citep{2015A&A...573A..40D, 
2018arXiv180403030L} 
to process the data. 
This includes image restoration with the multi-object multi-frame blind deconvolution 
\citep[MOMFBD, ][]{2005SoPh..228..191V} 
method and the procedure for consistency across narrowband channels of 
\cite{2012A&A...548A.114H}. 
High image quality was further aided with the SST adaptive optics system
\citep{2003SPIE.4853..370S} 
which has a 85-electrode deformable mirror operating at 2~kHz.

The data recorded during the best seeing conditions was observed on 22 September 2017. 
During the best periods, the Fried's parameter $r_0$ was above 50~cm, with a maximum of 79~cm 
\citep[for a discussion of measurements of $r_0$ by the SST adaptive optics system, see][]{2019A&A...626A..55S}. 
Unfortunately, the seeing was not consistently of high quality and the data set is not optimal for temporal evolution studies. 
Most of the analysis and data presented in Figs.~1--5 
is based on the CHROMIS and CRISP spectral scans recorded at 10:00:48~UT. 
The target area was the main sunspot in AR12681 at $(X,Y)=(-749\arcsec,-296\arcsec)$, $\mu=\cos \theta=0.54$ with $\theta$ the observing angle.
With CHROMIS, we sampled the \Hbeta\ line at 32 positions between $\pm$1.37~\AA\ with equidistant steps of 0.074~\AA\ around the line core and sparser in the wings to avoid line blends.
The time to complete a full \Hbeta\ scan was 11.1~s.
The CHROMIS data has a pixel scale of 0\farcs038 and the telescope diffraction limit ($\lambda/D$) is 0\farcs1 at $\lambda=4861$~\AA.
The CHROMIS instrument has an auxiliary wide-band (WB) channel that is equipped with a continuum filter which is centered at 4846~\AA\ and has a full-width at half-maximum (FWHM) of the transmission profile of 6.5~\AA.
This filter covers a spectral region that is dominated by continuum and has relatively weak spectral lines \citep[see][for a plot of the transmission profile in comparison with an atlas spectrum]{2018arXiv180403030L}. 

With CRISP, we sampled the \Halpha\ line at 32 positions between $\pm$1.85~\AA\ from the line core with equidistant steps of 0.1~\AA\ between $-1.6$ and $+1.3$~\AA.
In addition, CRISP was sampling the \ion{Fe}{i}~6301 and 6302~\AA\ line pair in spectropolarimetric mode, with 9 positions in \ion{Fe}{i}~6301 and 6 positions in \ion{Fe}{i}~6302, avoiding the telluric blend in the red wing. Furthermore, a continuum position was sampled between the two lines.
The time to complete full scans of the \Halpha\ and \ion{Fe}{i} spectral lines was 19.1~s. 
The pixel scale of the CRISP data is 0\farcs058.

The other dataset that we analyzed in detail was observed on 29 April 2016 and was centered on the main sunspot in AR12533 at $(X,Y)=(623\arcsec,8\arcsec)$, $\mu=0.75$.
The seeing conditions were very good for the whole 1~h 30~m duration of the time series which started at 09:43:09 UT. 
The $r_0$ values were averaging at about 20~cm with peaks up to 30~cm. 
The online material includes movies of the temporal evolution of the sunspot. For these movies we have applied frame selection by rejecting 32 low quality images, this corresponds to 12\% of the total of 267 time steps. 
The CRISP instrument was running a program with \ion{Ca}{ii}~8542~\AA\ spectropolarimetry and \Halpha\ imaging spectrometry at a cadence of 20~s. 
The \Halpha\ line was sampled at 15 positions between $\pm$1.5~\AA\ with 0.2~\AA\ steps between $\pm$1.2~\AA.
We compare \Halpha\ wing images with images from the CRISP 8542~\AA\ WB channel. 
For CRISP, the WB channel branches off after the prefilter so that contrary to CHROMIS, one cannot have imaging in a clean continuum band. 
The prefilter has a FWHM of 9.3~\AA\ and is centered on the \ion{Ca}{ii}~8542~\AA\ line.
The \ion{Ca}{ii}~8542~\AA\ spectra were not included in our analysis. 
This data was earlier analyzed by
\citet{2020A&A...638A..63D} 
to study penumbral micro jets and the co-aligned SST and IRIS data were publicly released as described by
\citet{2020A&A...641A.146R}. 

For the exploration of all data, verification of detected events, and the study and measurement of the dynamical evolution of PEBs in the 29 April 2016 data, we made use of CRISPEX
\citep{2012ApJ...750...22V}, 
a widget-based graphical user interface for exploration of multi-dimensional data sets written in the Interactive Data Language (IDL).

\section{Methods}
\label{sec:methods}

\paragraph{Inversions.}
We have performed Milne-Eddington (ME) inversions of the \ion{Fe}{i} line pair observed on 22 September 2017 to infer the magnetic field vector utilizing a parallel C++/Python implementation\footnote{\url{https://github.com/jaimedelacruz/pyMilne}} \citep{2019A&A...631A.153D}.
The magnetic field vector retrieved from the ME inversions are an average over the formation height of the \ion{Fe}{i} line pair.    
For these lines, the response of Stokes profiles to magnetic field is maximum around optical depth 0.1 at 5000~\AA\ in sunspot penumbrae \citep[e.g., see Fig.~9 of][]{2017A&A...599A..35J}.

We resolved the 180\degr\ ambiguity in our magnetic field vector measurements using the acute angle method \citep{1985tphr.conf..313S,1992A&A...265..296C}.
The inferred magnetic field vector in the line-of-sight frame of reference is projected to the disk center coordinates where $B_z$ represents the magnetic field component normal to the solar surface and $B_x$ and $B_y$ are the two orthogonal components projected onto the solar surface. 

To better resolve opposite polarity patches in the penumbra, we corrected for stray light prior to the inversions. We assumed a Gaussian point spread function (PSF) with FWHM of 1\farcs2 and 45\% stray light, following similar stray light corrections that were considered for CRISP/SST observations in earlier studies.
For example, \citet{2011Sci...333..316S} and 
\citet{2012A&A...540A..19S} 
compensated for stray light using a PSF with FWHM of 1\farcs2 and 56\% stray light contribution. 
\citet{2011ApJ...734L..18J} assumed 35\% stray light and a Gaussian PSF with FWHM of 1\farcs6.  
Moreover, from a detailed analysis of solar granulation contrast, 
\citet{2019A&A...626A..55S} 
concluded that stray light at the SST comes mainly from small-angle scattering and that the wings of the uncorrected PSF do not extend beyond 2\arcsec. 

\paragraph{\textit{k}-means clustering.}
We used the $k$-means clustering technique \citep{everitt_1972} to identify EB spectra in the \Hbeta{} spectral line observed on 22 September 2017. 
The $k$-means method is widely used for the characterization of a variety of solar phenomena and observations.
Examples include the classification of \MgH{} and k line profiles observed with \IRIS{} 
\citep{2019ApJ...875L..18S}, 
the identification of \MgH{} and k spectra in flares 
\citep{2018ApJ...861...62P}, 
and \CaK{} observations of on-disk spicules 
\citep{2019A&A...631L...5B,2021arXiv210107829B}. 
Our approach for clustering the \Hbeta{} spectra is very similar to that employed by 
\citet{2020A&A...641L...5J}  
and \citet{2020A&A_joshi}  
to identify QSEBs in their \Hbeta{} observations.
With the $k$-means method we divided \Hbeta{} spectra into 100 clusters and each cluster is represented by the mean of all profiles in that cluster. This mean profile is referred to as representative profile (RP). 
Out of 100~RPs we found that 29~RPs show line wing enhancement that is characteristic of EBs. Of these 29 selected RPs with enhanced wings, 25 RPs have essentially an unaffected line core, while the rest show an intensity enhancement in the line core.
Inclusion of the four RPs, that have an enhanced line core as EB profiles, is motivated by 
\citet{2020A&A...641L...5J}  
who found that unlike typical \Halpha\ EB profiles, some EBs can show raised intensity level even in the \Hbeta\ line core.
A detailed description of selected RPs with EB-like \Hbeta{} spectral profiles is provided in Sect.~\ref{sec:results}.   

Based on spatial locations of selected RPs, we created a binary mask which was then used to perform two dimensional (2D) connected component labeling \citep{labeling_1996} that assigns a unique label for each isolated patch in the binary mask. 
We then used the labels to estimate their area, brightness enhancement and radial distance from the geometric center of the sunspot for each individual EB.   
A detailed statistical analysis of these parameters is presented in Sect.~\ref{sec:results}.


\section{Results}
\label{sec:results}

\subsection{PEB morphology and general appearance}
\label{sec:sub_morphology}

\begin{figure*}[!ht]
\includegraphics[width=\textwidth]{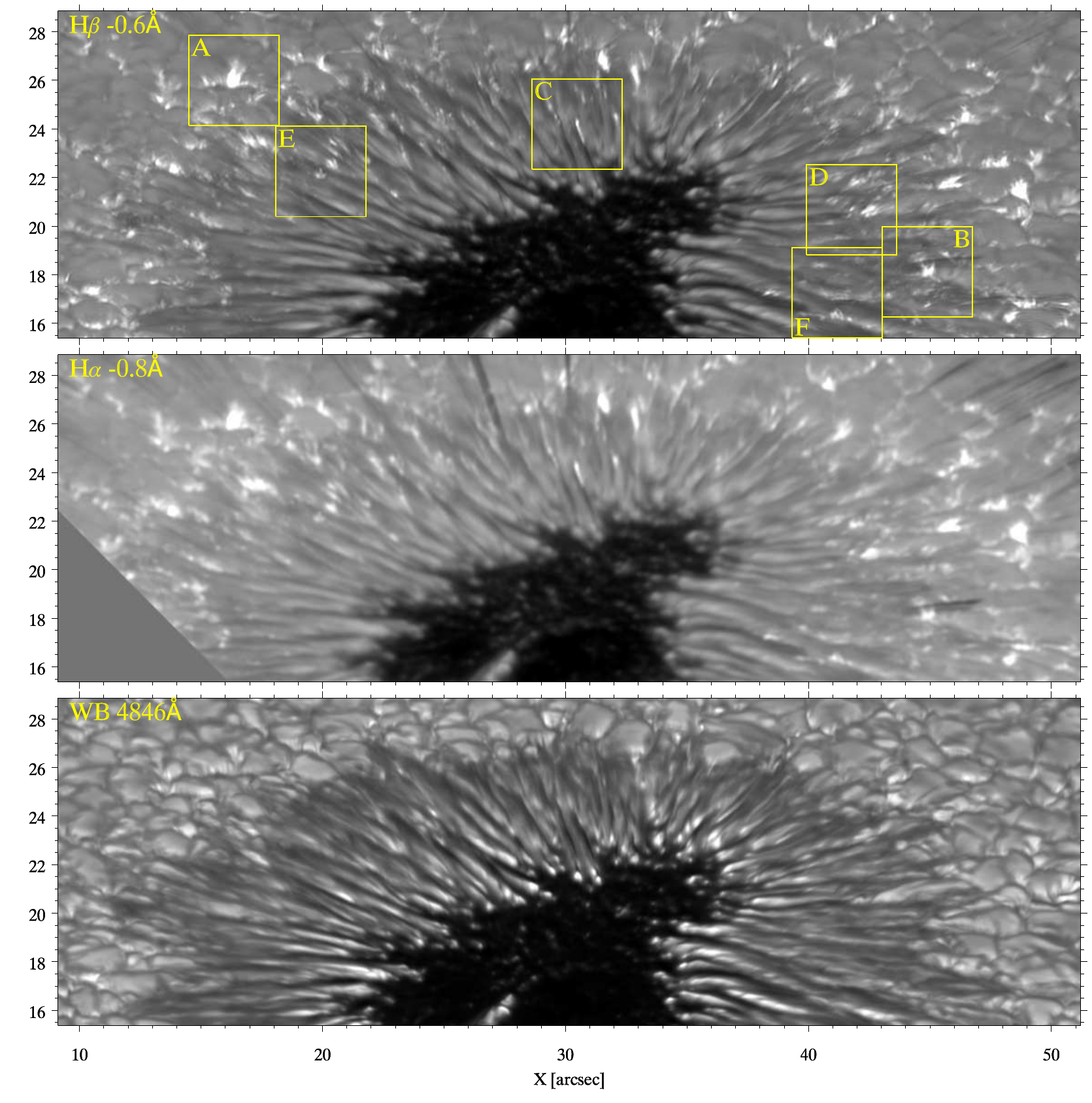}
\caption{\label{fig:overview}%
Limb-side part of the sunspot in AR12681 observed on 22 September 2017
in \Hbeta\ and \Halpha\ blue wing and CHROMIS WB 4846~\AA. PEBs are
visible as small bright features all over the penumbra, some with
clear flame morphology sticking straight up between filaments. These
PEBs are invisible in the continuum WB image. The direction to the
nearest limb is approximately upward along the $y$-axis. The top image
includes six squares labelled A--F that mark ROIs that are shown in
detail in Fig.~\ref{fig:details}. An animation of this figure is available as online
material at \url{https://www.mn.uio.no/astro/english/people/aca/rouppe/movies/}. 
This animation shows a spectral scan through the \Hbeta\ and \Halpha\ lines. 
}
\end{figure*}

\begin{figure*}[!ht]
\includegraphics[width=0.49\textwidth]{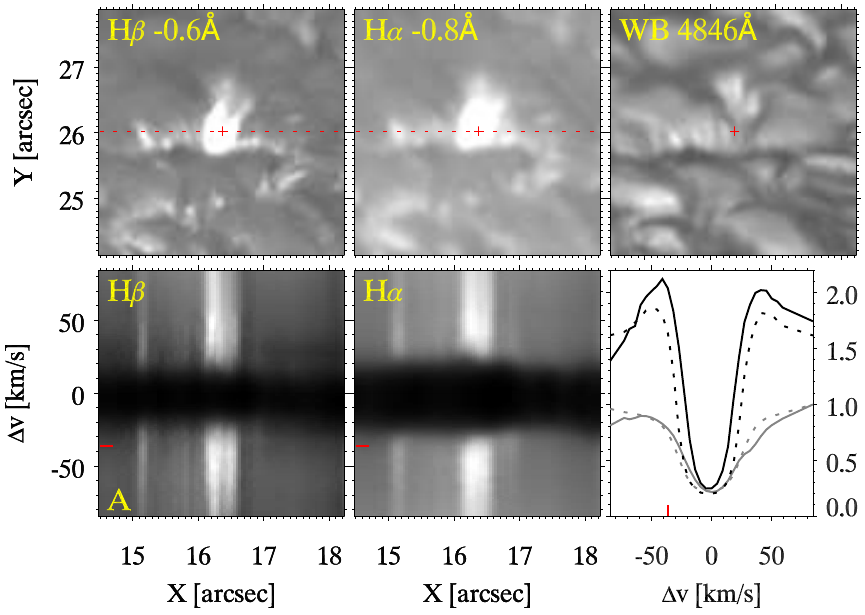}
\includegraphics[width=0.49\textwidth]{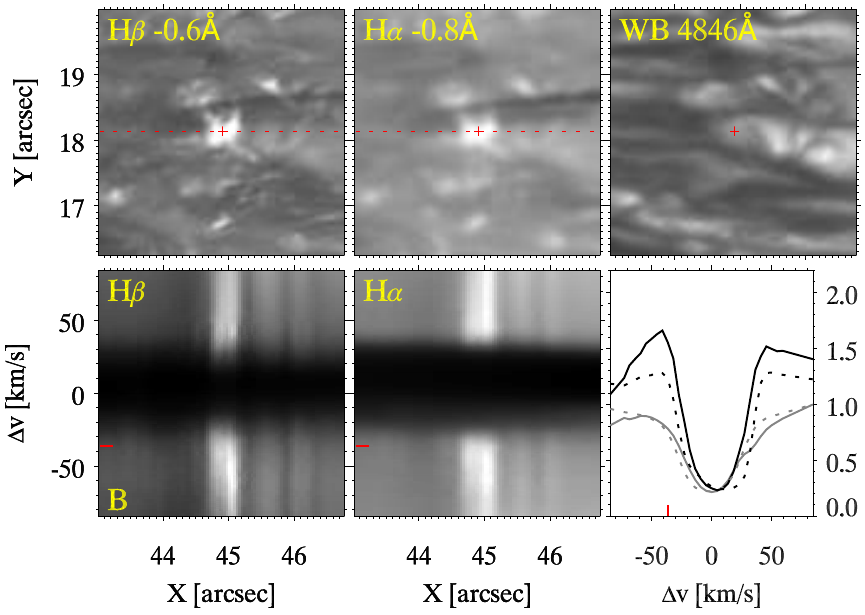} \\
\includegraphics[width=0.49\textwidth]{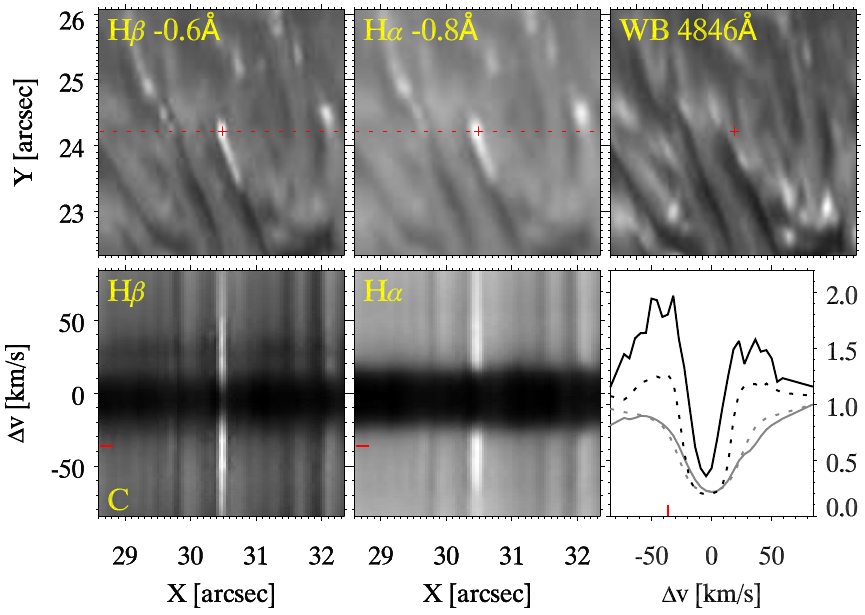}
\includegraphics[width=0.49\textwidth]{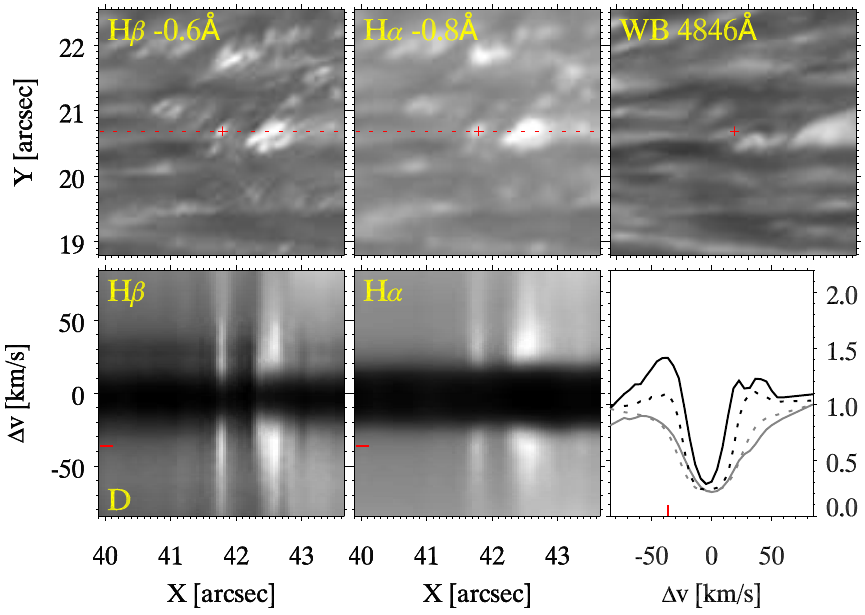} \\
\includegraphics[width=0.49\textwidth]{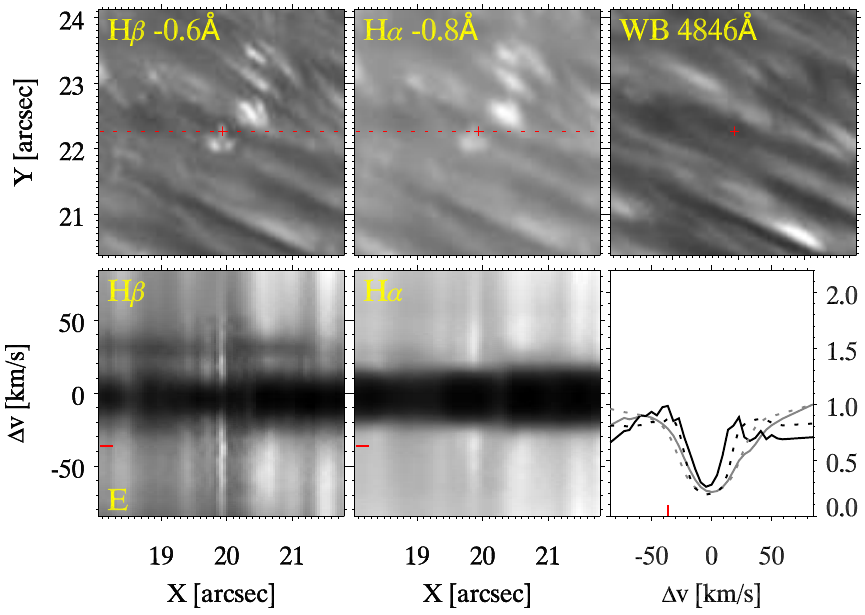}
\includegraphics[width=0.49\textwidth]{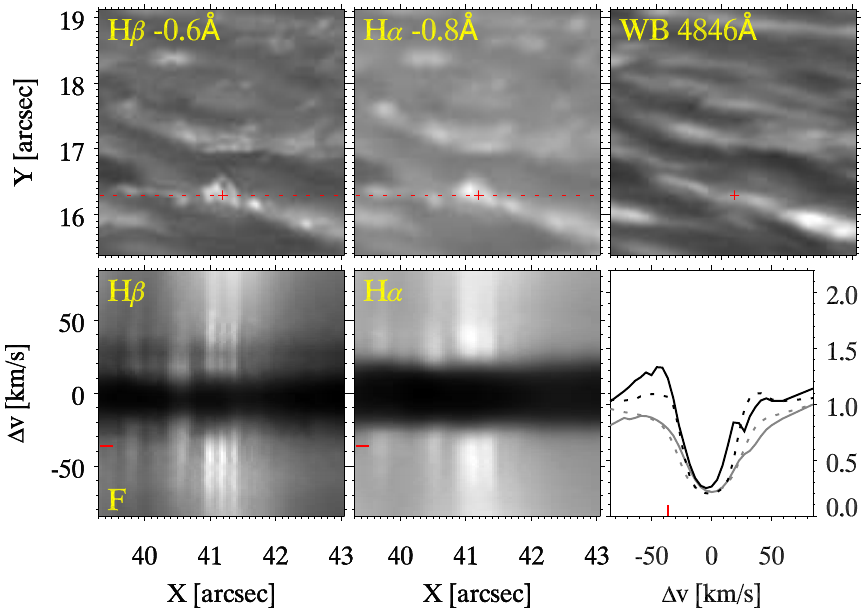} \\
\caption{\label{fig:details}%
Details of EBs in and outside the penumbra of the sunspot shown in Fig.~\ref{fig:overview} in 6 ROIs. The spatial $X,Y$ coordinates are at the same scale as Fig.~\ref{fig:overview}. The top row of panels for each ROI shows \Hbeta\ and \Halpha\ blue wing and CHROMIS WB 4846~\AA\ images. The bottom left panels show $\lambda x$-diagrams of the spectral profiles along the red dotted line in the panels above. The bottom right panel shows spectral profiles for \Hbeta\ (solid black line) and \Halpha\ (dashed line) from the position of the red cross in the top left panels. The thin gray profiles are reference spectral profiles averaged over an area outside the sunspot. The intensity scaling is normalized to the level of the far red wing of the reference profile. The red tickmark in the bottom row panels indicates the line position of the wing images in the top left. ROI~A is centered on a strong EB outside the sunspot. ROI~B is centered on an EB at the outer edge of the penumbra. All other examples are PEBs inside the penumbra.
}
\end{figure*}

Figure~\ref{fig:overview} shows the limb-side part of the 22 September 2017 sunspot in the blue wings of \Hbeta\ and \Halpha\ as well as in CHROMIS WB 4846~\AA. 
The offset from line core was chosen to be close to the maximum of the typical EB profile as to show EBs at highest contrast. 
Some prominent EBs are visible as pronounced flames in the moat around the sunspot outside the penumbra. 
As expected, the EBs are not visible in the continuum dominated WB image. 
Inside the penumbra, there are a large number of small bright features present in the Balmer wing images but clearest in the \Hbeta\ wing image and not visible in the WB image. 
Some of these appear as small linear features sticking straight up from between the penumbral filaments, resembling the larger EB flames in the surrounding sunspot moat. 

The animation associated with Fig.~\ref{fig:overview} shows a spectral line scan through the \Hbeta{} and \Halpha{} lines for comparison. 
From the animation it is evident that in the penumbra the EB-like brightenings in the \Hbeta{} wings also persist in and close to the line core wavelength positions.      
However, these compact brightenings in the \Hbeta{} line core are absent in the \Halpha{} line core which predominantly show chromospheric superpenumbral fibril structures.    

Figure~\ref{fig:details} zooms in on 6 regions of interest (ROI). 
In the upper left, ROI~A is centered on the most prominent EB in the FOV, with the telltale flame towering about 600~km above the intergranular lane from where it appears to emanate. 
The CHROMIS WB image shows no trace of the EB, only some striations in the background faculae, which are unrelated to the EB phenomenon. 
The $\lambda x$-diagrams and spectral panel show the well-known characteristic EB Balmer profile with enhanced wings and unaffected line core. 
The peak wing enhancement is more than 2 times the level of the reference profile which is averaged over a quiet region. 
The higher contrast and higher spatial resolution in the \Hbeta\ data compared to \Halpha\ is clear, for example from the fine structure and spatial variation in the $\lambda x$-diagram.
The EB in ROI~A serves as reference for the EBs presented in the other ROIs.

In ROI~B, a clear EB flame is located at the outer edge of the penumbra. 
The vertical extension of this flame has a length of about 450~km. 
The other four ROIs are all inside the penumbra and are centered on PEBs. 
%
Of these, ROI~F is centered on the tallest flame which has a length of about 350~km. The \Hbeta\ wing image shows clear substructure in the PEB while it is more an extended fuzzy feature in the \Halpha\ wing image. For this case, the wing enhancement in the \Hbeta\ profile is only slightly larger than in \Halpha.
For the PEB examples in ROIs~C and D, the differences in wing enhancement is larger, in particular in ROI~C where the peak in wing enhancement is almost as high as for the large EB in ROI~A. 
Flame morphology in ROI~C might be difficult to discern because the PEB is aligned along the penumbral filaments that, in this part of the penumbra, are aligned in the direction of the nearest limb (i.e. along the line-of-sight).
ROI~E is centered on a PEB that has hardly enhanced wings in the profile plot but clearly shows a little flame in the \Hbeta\ wing image and is unmistakably present in the $\lambda x$-diagram. While this PEB might be weak, its absence in the WB image is striking.
This weak event is detected as a PEB with the $k$-means method. 

In all of these ROIs the penumbra in WB appears smoother than in the Balmer wing images. Particularly in the \Hbeta\ wing there are many small bright features, like bright ``crumbs'', scattered over the penumbra. Some of these are very bright and show the characteristic EB wing enhancement and are clear PEBs. Many others show only subtle wing enhancement but are notably absent in the WB image. 
To the left of the PEB, in ROI~F, the red dashed line crosses some of these ``crumbs'' and the $\lambda x$-diagram shows wing enhancement when compared to their surroundings, but clearly not so much as the central PEB.

\begin{figure*}[!ht]
\includegraphics[width=\textwidth]{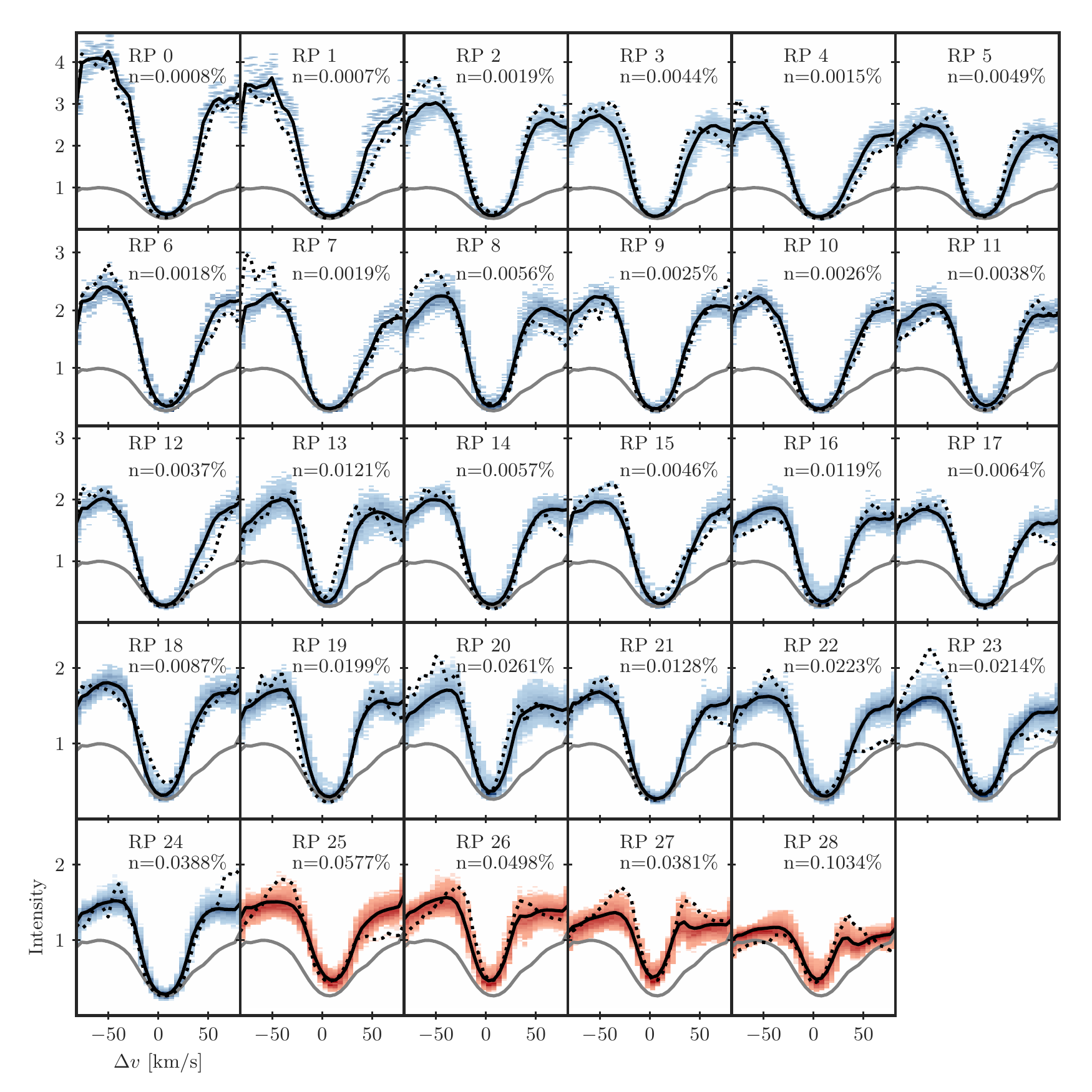}
\caption{\label{fig:RPs}%
Twenty nine representative profiles (RPs) from the $k$-means clustering of the \Hbeta{} line that are identified as signature of EB. The black lines show RPs whereas shaded colored areas represent density distribution of \Hbeta{} spectra within a cluster; darker shades indicate higher density. Within a particular cluster, the \Hbeta{} profile that is farthest (measured in euclidean distance) from the corresponding RPs is shown by the black dotted line. As reference, the average quiet Sun profile (gray line) is plotted in each panel. RPs 0--24 show the typical EB-like \Hbeta{} profiles, i.e., enhanced wings and unaffected line core, while RPs 25--28 display both an enhancement in the wings as well as in the line core. The parameter $n$ represents the number of pixels in a cluster as percentage of the total of $\sim1.73\times10^6$ pixels.    
}
\end{figure*}

Figure~\ref{fig:RPs} shows all \Hbeta{} RPs that have been identified as showing EB spectral signatures. 
Representative profiles 0--24 have profiles similar to typical \Halpha\ EB profiles with enhanced wings and essentially unaffected line core whereas RPs 25--28 display intensity enhancement in the line core along with enhancement in the wings. 
Each detected EB in our dataset displays a combination of RPs plotted in Fig.~\ref{fig:RPs}.  
For example, the PEB shown in the ROI~E of Fig.~\ref{fig:details} is identified as a line core brightening and represented by a combination of RPs 27 and 28.
Similarly, a part of the PEB in ROI~C is identified as a line core brightening by RP 25. 
The rest of the EBs and PEBs in Fig.~\ref{fig:details} predominantly exhibit wing intensity enhancement in combination with unaffected line cores and are clustered following RP 0--24.      

Besides the RPs, Fig.~\ref{fig:RPs} shows a density distribution of all \Hbeta\ profiles that are included in each cluster. 
The density distributions are narrow and centered around the RPs. However, in some clusters the farthest profile shows some significant deviation from the corresponding RP.  
For example, in clusters represented by RP 7 and 23, the farthest profiles have quite different shape as compared to their respective RPs. 
Nevertheless, these farthest profiles also show characteristic EB-like spectral profiles. 

\subsection{Magnetic field environment} 

\begin{figure*}[!ht]
\includegraphics[width=\textwidth]{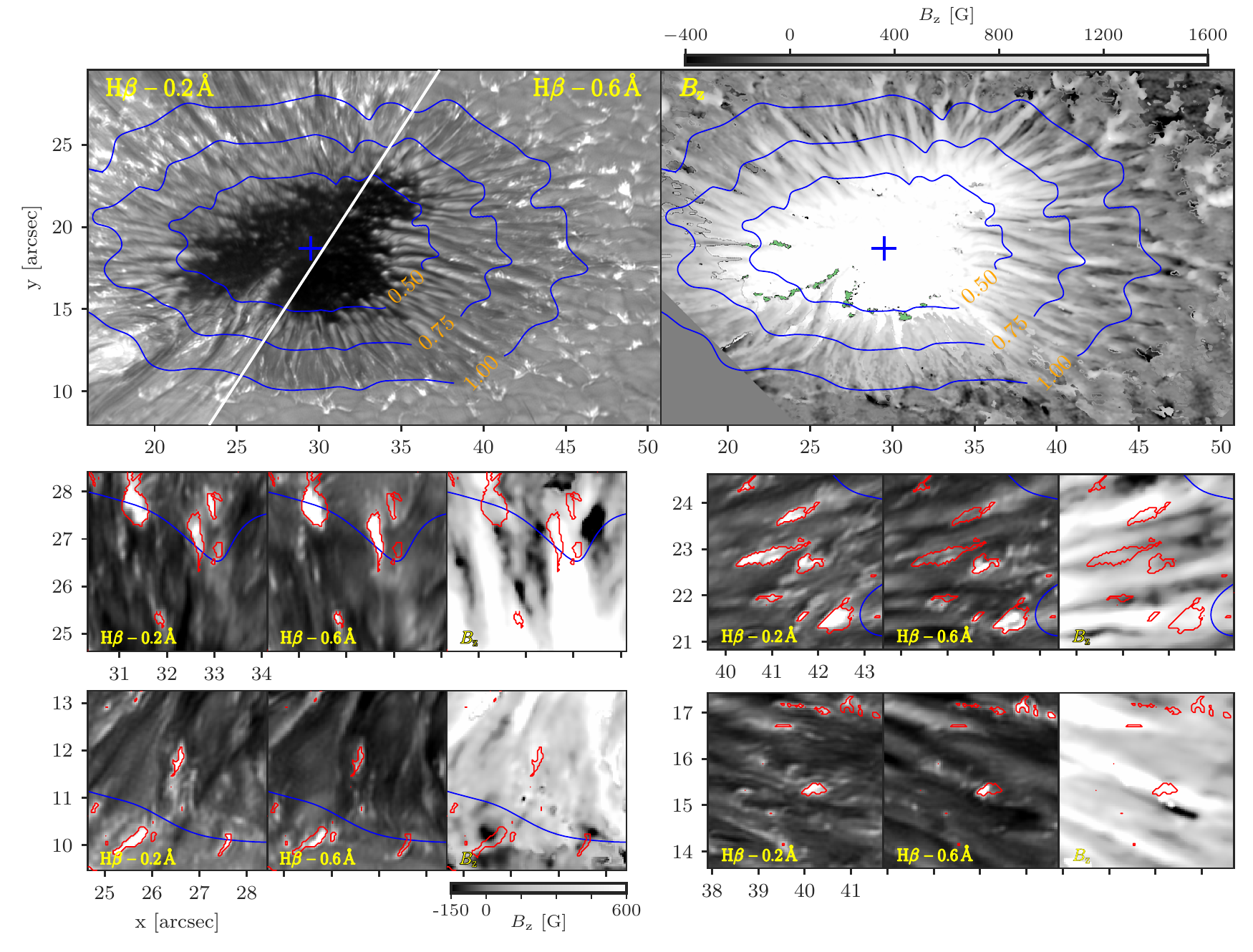}
\caption{\label{fig:Bz_map}%
The location of PEBs compared to the vertical magnetic field $B_z$. The top left panel shows a split image of the sunspot observed on 22 September 2017 with the left part in the \Hbeta\ blue wing at $-0.2$~\AA\ offset and the right part at $-0.6$~\AA. 
The blue contours indicate the radial distance $r/\mathrm{R_{spot}}$
to the umbral center that is marked with the blue cross. The contour
for $r/\mathrm{R_{spot}}=1.00$ is the outer penumbra boundary, defined
from the associated WB image. The top right panel shows the $B_z$ map, derived from ME inversions of the \ion{Fe}{i} lines, scaled between $-400$ and $+1600$~G. Regions with artifacts due to the de-projection method are marked in green. The sets of panels at the bottom show four ROIs in \Hbeta\ $-0.2$~\AA, $-0.6$~\AA, and $B_z$ respectively. Red contours outline PEBs detected through the $k$-means method.    
}
\end{figure*}


\begin{figure}[!ht]
\includegraphics[width=0.48\textwidth]{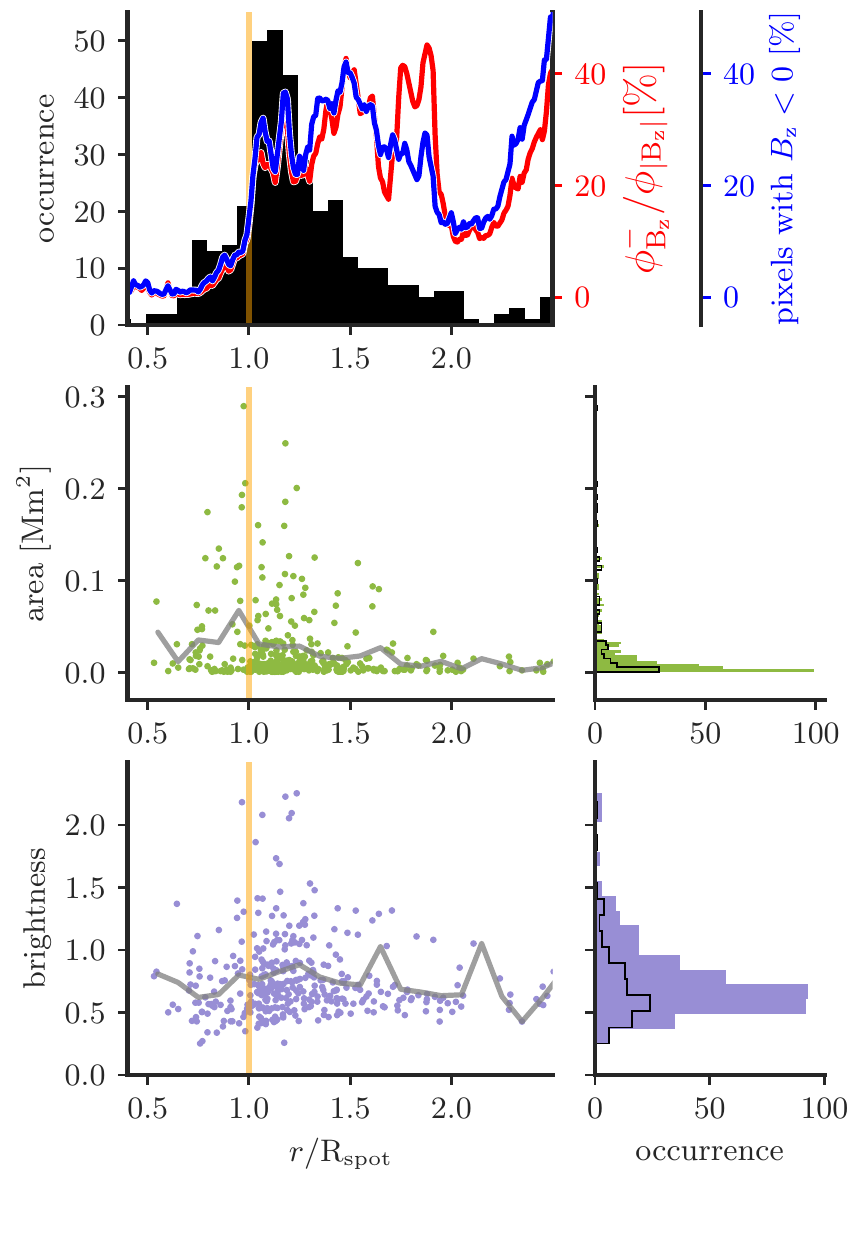}
\caption{\label{fig:radial}%
Distribution of EBs and their properties with respect to radial distance from the sunspot center (observed on 22 September 2017). The outer sunspot boundary is at $r/\mathrm{R}_\mathrm{spot}=1$ and is marked with the yellow vertical line, also see the contours in Fig.~\ref{fig:Bz_map}.
The statistics are based on $k$-means detections, the total number of
EB detections is 372, of which 108 are PEBs. The top panel shows the
EB occurrence. The red curve shows the ratio of negative magnetic
polarity flux relative to total absolute flux (the sunspot is
dominated by positive polarity). The blue curve shows the fraction of
pixels with negative polarity relative to all pixels with significant
magnetic signal ($|B_z| > 50$~G). The middle panels show the area of
the EB detections. The bottom panels show the \Hbeta\ wing brightness
enhancement of the brightest pixel in the EB detection relative to the
local background in a 100$\times$100 pixel area and excluding EB
detection pixels. The brightness enhancement is relative to the outer
most wavelength positions on both sides of the line center of the
background \Hbeta\ profile and on a scaling set by the normalized
quiet Sun reference profile. For the two bottom rows, the right panels show occurrence histograms with the black line outlining the histograms for PEBs. 
The histogram bin size is 0.003~Mm$^2$ in area and 0.12 in brightness enhancement. 
The grey lines in the left panels mark the average values for each radial distance $r/\mathrm{R}_\mathrm{spot}$.}
\end{figure}

In order to study the occurrence of PEBs with respect to the magnetic field in the vicinity, we compare EB detections from the $k$-means method with the $B_z$ map derived from the \ion{Fe}{i} lines. 
This is illustrated in Fig.~\ref{fig:Bz_map}. 
The sunspot is dominated by positive magnetic polarity but the $B_z$ map also shows many small isolated patches with significant opposite (negative) polarity within the outer penumbra boundary. 
We find that many PEBs are located in the vicinity of these opposite polarity patches. 
This can be seen at close look in the four ROIs in the bottom of Fig.~\ref{fig:Bz_map}.
Red contours outline EB detections and  there are some clear examples
of PEBs that are located at or close to the interface where opposite polarities meet. 
We note however, that we also find PEBs located in unipolar regions, for example the PEB in the center of the lower left ROI. 

As mentioned before, PEB brightenings can also be visible in and close to the \Hbeta{} line core, see the left \Hbeta\ $-0.2$~\AA\ part of Fig.~\ref{fig:Bz_map} (top left), that displays numerous compact brightenings in the penumbra.
A number of PEBs that can be seen in the \Hbeta{} line core are shown in more detail in the ROIs presented in the bottom of Fig.~\ref{fig:Bz_map}.  
For example, in the upper-left ROI, two big PEBs at the sunspot boundary are visible in the wing as well as close to the line core.   
The PEBs at $(X,Y)=(41\farcs5,23\farcs6)$ and $(X,Y)=(41\farcs0,22\farcs8)$ in the upper-right ROI are predominantly visible at $-0.2$~\AA\, while they have only subtle brightenings in the outer \Hbeta{} line wing. 

The statistics shown in Fig.~\ref{fig:radial} provide a quantified context of the observation that PEBs are often found in the vicinity of opposite polarities: the top diagram shows that both the number of PEBs and the contribution from opposite polarity patches increase towards the outer penumbra. 
Both the relative area (blue curve) and opposite polarity flux (red curve) increase to more than 10\% at the outer penumbra boundary. 

With the $k$-means clustering method, we detected a total of 372 EBs of which 108 are in the penumbra. 
We found no EBs in the umbra. 
In the inner penumbra, $0.5 \le r/\mathrm{R}_\mathrm{spot} \le 0.75$, the number density of detected PEBs is 0.29~Mm$^{-2}$, and the fraction of the total area covered by PEBs is 0.007 (i.e., the area filling factor).
In the outer penumbra, $0.75 < r/\mathrm{R}_\mathrm{spot} \le 1$, the number density is 0.76~Mm$^{-2}$, and the area filling factor 0.032.
In the immediate surroundings of the sunspot, in the moat, $1 < r/\mathrm{R}_\mathrm{spot} \le 1.25$, the EB number density is 1.72~Mm$^{-2}$, and the area filling factor 0.037.
%
The number density of all 372 EBs detected over the full CHROMIS FOV is 0.27~Mm$^{-2}$.
%
%
For two other \Hbeta\ spectral scans of this sunspot, recorded under less optimal seeing conditions, we find fewer but comparable numbers of EB detections: for a scan with quiet Sun granulation contrast 15.0\%, there are 304 EB detections of which 90 PEBs, and for a scan with 14.6\% contrast, there are 252 EBs of which 75 PEB (the contrast for the best scan is 15.7\%). So about 30\% of the EBs detected in the FOV are inside the penumbra.  

%

Figure~\ref{fig:radial} further provides statistics on the area and brightness enhancement of the EB detections. 
The largest PEBs are found towards the outer penumbra and PEBs do not stand out as being smaller or larger than EBs outside the sunspot.
The mean area for PEBs is 0.039~Mm$^2$ (standard deviation $\sigma=0.055$~Mm$^2$) and for EBs outside the sunspot 0.022~Mm$^2$ ($\sigma=0.041$~Mm$^2$).
%
The trend in the area distribution has a sharp cut off at 0.0037~Mm$^2$ (five pixels) which is set by the spatial resolution.
This suggests that there exist smaller EBs that are not resolved. 
Many small bright features in the \Hbeta{} wings described as ``crumbs'' in Sect.~\ref{sec:sub_morphology} were not detected by the $k$-means method.
In some cases where these features were detected, only a few of the brightest pixels were identified as PEBs and not the whole morphological structure. Thus, these detections also contribute to the population of PEBs with smallest areas.
We excluded all EB events with area less than five pixels in our statistical analysis.   

Also in terms of wing brightness enhancement, as shown in the bottom of Fig.~\ref{fig:radial}, PEBs do not stand out as compared to EBs.
The average wing brightness enhancement for PEBs is 0.72 ($\sigma=0.33$) and for EBs outside the sunspot 0.78 ($\sigma=0.35$).
Here, the \Hbeta{} wing enhancement was measured against the average intensity of the outermost wavelength positions in the local background (over an area of $100\times100$ pixels).
The majority of the EBs, within the penumbra as well as in the surroundings of the sunspot, have a brightness enhancement between 0.5 and 1. 
However, some PEBs in the outer penumbra and some EBs in close proximity of the sunspot are brighter, and for some, the intensity enhancement relative to the local surroundings is larger than 2.
The brightest EBs were classified as RP~0--2 (see Fig.~\ref{fig:RPs}) for which the blue wing is raised more than three times the level of the reference quiet Sun.

\subsection{Temporal evolution}

\begin{figure*}[!ht]
\includegraphics[width=\textwidth]{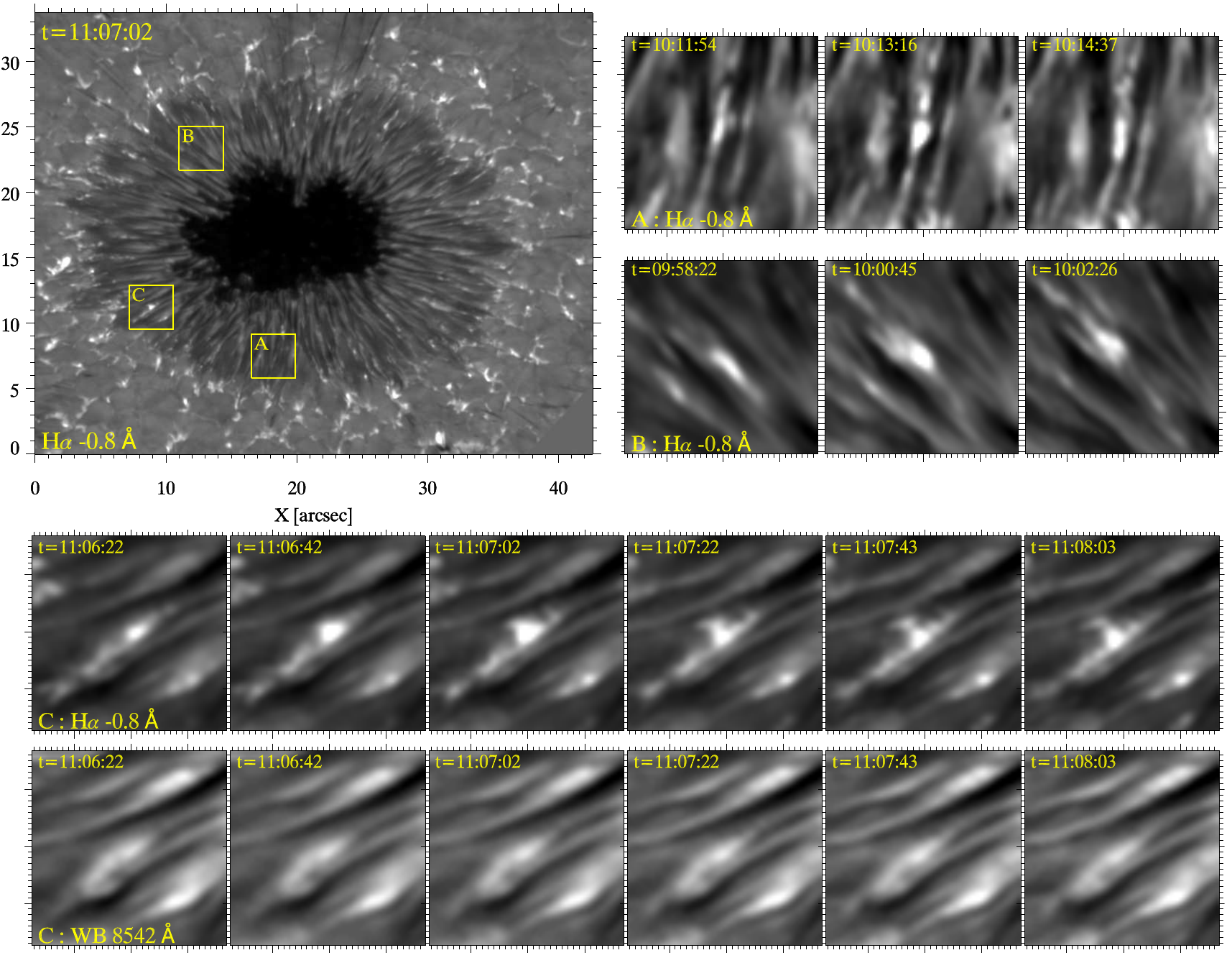}
\caption{\label{fig:evolution}%
Temporal evolution of PEBs in the sunspot in AR12533 observed on 29 April 2016. The top left image shows an \Halpha\ blue wing image at $-0.8$~\AA\ with three regions of interest (ROI) marked with labels A, B, and C. The temporal evolution for these ROIs is shown in the rows with smaller \Halpha\ wing images where the time is marked in the top left. The bottom row of images shows the temporal evolution in ROI~C in WB~8542~\AA\ for comparison. The spacing between large tick marks in the ROI images is 1\arcsec. The contrast in the top left overview image is enhanced by applying a gamma correction with $\Gamma$=2, all other images have linear scaling on a common scale for each ROI. Three animations associated to this figure are available as online material: a movie of the sunspot in the \Halpha\ blue wing as in the upper left panel, the corresponding movie in WB~8542~\AA, and a combined movie showing the left part of the sunspot. See \url{https://www.mn.uio.no/astro/english/people/aca/rouppe/movies/}.
}
\end{figure*}

The temporal evolution of PEBs was studied in the 29 April 2016 observations of the sunspot in AR12533, see Fig.~\ref{fig:evolution}. 
The online material includes a movie of the full 90~min sequence of the entire sunspot at $-0.8$~\AA\ offset from \Halpha\ line center, equivalent to the FOV shown in the upper left panel. 
The movie shows many small bright PEBs in the penumbra that generally move radially outward, away from the umbra. These are not visible in the reference WB~8542~\AA\ movie. This difference in visibility is best seen in the 3rd movie that zooms in on the left part of the penumbra and combines the \Halpha\ blue wing and WB~8542~\AA, as well as a panel that blinks between these diagnostics. 
There are numerous examples of PEBs that originate in the penumbra, migrate outwards and eventually cross the outer penumbra boundary where they continue their outward migration in the sunspot moat flow.
From inspection of the \Halpha\ blue wing movie, it is clear that most of the PEBs are found in the outer regions of the penumbra. This is similar to what is described above for the 22 September 2017 sunspot and what was found from the $k$-means detections. 

We tracked 32 events to measure lifetimes, trajectories and velocities. These PEBs were selected on the basis of clear visibility throughout their lifetime and regarded as a representative sample.
We measured PEB lifetimes ranging between 1 and 9~min, and an average lifetime of about 3~min. During their lifetime, these PEBs traveled distances ranging between 100 and 1640~km, with an average of about 650~km. They traveled at an average speed of 3.7~\kms\ and the maximum speed measured is almost 13~\kms. 
These velocities are apparent motions and from these observations we cannot determine whether these are real plasma flows or result from a moving front of for example reconnection moving along a magnetic interface.

Figure~\ref{fig:evolution} shows the evolution of selected PEBs in sequences of small images for three ROIs. 
The three images for ROI~A cover 2:43~min during which the PEB migrates over a distance of 340~km with an average speed of 2~\kms. The PEB flares up for a duration of 102~s, with its brightest moment at 10:13:16~UT in the middle panel. 
The PEB in ROI~B migrates over 365~km with an average speed of 1.4~\kms. During its lifetime, the PEB splits into a number of bright substructures, this is visible in the 3rd panel. 
The sequence for ROI~C covers 1:41~min of a total lifetime of the PEB of 5:04~min. This PEB shows a clear flame structure which is strikingly absent in the reference row of WB~8542~\AA\ images. This flame appears to eject a small bright blob that is visible in the three last panels. This ejection can be followed for 2:22~min during which it moves at a maximum speed of almost 11~\kms. The PEB itself appears to move at a maximum speed of almost 4~\kms\ while it moves at an average speed of 2~\kms\ during a migration over 650~km. 

The rapid variability we see for these selected examples and other PEBs in the time series is clearly limited by the temporal resolution. There often are significant variations in brightness and morphology (e.g., in the form of splitting, merging, and ejections) between subsequent time steps. This suggests that PEBs are changing on a shorter time scale than 20~s. 

\section{Discussion and conclusion}

\label{sec:discussion}

Using high spatial resolution observations in the Balmer \Halpha\ and \Hbeta\ lines, we find large numbers of EBs in the penumbrae of sunspots. 
The EB nature of these penumbral events is established by 1: characteristic spectral profiles with often strongly enhanced wings, 2: flame morphology under slanted viewing angle, 3: rapid temporal variability in brightness and morphology, and 4: absence in concurrent continuum passbands. 

We find many small patches in the penumbra with characteristic EB wing enhancement and note that there is considerable spread in the level of enhancement: some reach the level of strong EBs traditionally found in active region flux emergence regions with wing enhancement well above twice that of quiet Sun, others have weak wing enhancement that is only discernible in contrast to weak background penumbral profiles.  
In the \Hbeta\ line, we find that PEBs do not stand out in terms of area or wing brightness as compared to EBs in the surroundings of the sunspot. 
We do note however, that PEBs are easier to discern in \Hbeta\ than in \Halpha. The shorter wavelength of \Hbeta\ offers the advantage of higher spatial resolution and higher contrast. Furthermore, we observe that the sunspot in the \Halpha\ line is much more dominated by dense chromospheric fibrils. It appears that the sunspot chromosphere has much less opacity in \Hbeta. 
Recently, from non-LTE radiative transfer calculations, \citet{2020SCPMA..6319611Z} 
concluded that \Hbeta\ Stokes signals originate from the sunspot umbra photosphere. 
The difference between the \Halpha\ and \Hbeta\ lines is well illustrated by the line scan animation associated with Fig.~\ref{fig:overview} in the online material. At wing offsets that have highest EB contrast, the \Halpha\ line is much more dominated by the chromospheric superpenumbra fibrils than the \Hbeta\ line. These combined reasons make it more difficult to detect PEBs in \Halpha\ and appreciate the ubiquity of PEBs.
Here we present detailed analysis of two different sunspots, but we note that we observe large numbers of PEBs in at least 11 other sunspot datasets that we have acquired over the past ten years. 
In the 22 September 2017 dataset, we find more than 100 PEBs in the highest quality \Hbeta\ line scan which corresponds to almost 30\% of all detected EBs.
The number density of PEBs is higher than the average number density of EBs over the whole FOV. It is only in the sunspot moat, just outside the penumbra, that the number density of EBs is higher than in the outer penumbra. In the moat we detect on average about 3 EBs per typical granule, considering an average area of a granule of 1.75~Mm$^2$ 
\citep[see][]{2018LRSP...15....6R}. 
For the outer penumbra, we detect about 1.3 PEBs per typical granule area. 
%
In the quiet Sun, \citet{2020A&A...641L...5J} 
found a QSEB number density of 0.09~Mm$^{-2}$, which is more than 8 times lower than the number densities of PEBs (0.76~Mm$^{-2}$) and 19 times lower than EBs in the moat (1.72~Mm$^{-2}$).

Many events show clear flame morphology under inclined viewing angle which underlines the similarity with EBs in active regions and quiet Sun. 
The rapid variability and dynamics we see in the 29 April 2016 time series remind of the rapid variability found in EB flames in active regions 
\citep{2011ApJ...736...71W} 
and quiet Sun
\citep{2016A&A...592A.100R}. 
We note however, that the temporal cadence of these studies ($\sim1$~s) is much faster than for the time series presented here (20~s).

Establishing the ubiquity of EBs in the penumbra is aided by the concurrent continuum observations that are available through the CHROMIS WB channel. 
Absence of a bright feature in the associated continuum image confirms the Balmer line wing enhancement. 
The EB features are as absent in the penumbra as they are outside the sunspot and this further confirms the EB nature of PEBs.
The \Hbeta\ wing images show many small bright features that are absent in the WB image but have too weak wing enhancement to be (fully) detected by the $k$-means method (in Sect.~\ref{sec:sub_morphology} described as ``crumbs'').  This suggests that PEBs are more prevalent than the detection numbers from the $k$-means method suggest.

For CRISP \Halpha\ observations, a clean continuum channel is not as readily available as the CRISP WB channel shares the same prefilter as the CRISP narrow band images. 
CRISP WB~6563~\AA\ images show EBs because the CRISP prefilter transmission profile has relatively wide passband (FWHM=4.9~\AA) and is centered on the \Halpha\ line. 
For the 29 April 2016 time series (see Fig.~\ref{fig:evolution}) we compare \Halpha\ blue wing images with concurrent CRISP WB~8542~\AA\ images since EB signature in this channel is weaker due to wider passband and generally weaker EB emission in \ion{Ca}{ii}~8542~\AA.

The presence of EBs in the sunspot penumbra has been reported before. 
For example, a number of small EBs inside the penumbra of a small sunspot can be seen in the \Halpha\ wing detection maps of 
\citet{2013SoPh..283..307N},  
and \citet{2013ApJ...779..143R} 
report the observation of EB profiles in the \ion{Ca}{ii}~8542~\AA\ line for two events in a study of penumbral transients. 
However, this is the first time that the presence of large numbers of EBs in the penumbra is reported.

The significance of EBs lies in their capacity of being markers of magnetic reconnection in the low solar atmosphere. 
Numerical simulations demonstrated that enhanced Balmer wing emission and flame morphology stems from heating along current sheets at reconnection sites 
\citep{2017ApJ...839...22H, 
2019A&A...626A..33H, 
2017A&A...601A.122D}. 
The flames we observe for PEBs appear to be rooted deep down in the penumbra photosphere, in similar fashion as for EB flames in active regions and quiet Sun. 
Further support for PEBs being markers of magnetic reconnection in the deep penumbra photosphere comes from PEB detections being located in areas where opposite polarities are in close proximity.
The sunspot of 22 September 2017 is of positive magnetic polarity. The $B_z$ map (Fig.~\ref{fig:Bz_map}) reveals the presence of many isolated patches of opposite (negative) polarity within the penumbra. 
Many PEBs are located in the vicinity of these opposite polarity patches and some are located right at the interface where the two magnetic polarities meet.
We also observe that the number density of PEBs increases toward the outer penumbra, following the same trend of increasing opposite polarity flux with increasing distance from the sunspot umbra. 
We note however, that there are a few limitations that need to be kept in mind when combining the $B_z$ and EB detection maps for inferring that magnetic reconnection is taking place:
spectral line inversions are sensitive to a limited range in height and simplifications assumed for the ME inversion method imply uncertainties. 
We estimate that our ME inversions of the \ion{Fe}{i} lines are valid as $\vec{B}$ field measurements over a height range over a few 100~km in the upper photosphere
\citep[see][]{2010A&A...514A..91G}.
%
\citet{2017A&A...599A..35J}  
have shown that opposite polarity magnetic flux found in the deeper penumbra could be more than four times larger than that in the the middle and upper photosphere. 
Therefore, there is solid ground to believe that our ME inversions which provide height-independent magnetic field vectors are not able to resolve all opposite polarity patches in the penumbra.      
Furthermore, stray light makes it difficult to detect weak signals and adds to the uncertainty in the interpretation. We have applied a correction for stray light that is consistent with previous studies but the full impact of stray light on our measurements remains unknown.
Further uncertainties come from line-of-sight obscuration due to corrugation of the penumbral optical surface and it may be possible that regions with opposite polarity are hidden behind elevated foreground structures. 
Apart from these observational limitations that mitigate the detection of opposite polarity patches, it should be stressed that the condition of diametrically opposite direction fields is not strictly required for reconnection to take place. Even in areas that appear unipolar in observations, the complex magnetic topology of the penumbra can be expected to host gradients in the magnetic field that allow for small-angle magnetic reconnection.

The large number of PEBs we observe suggests that magnetic reconnection is a very frequently occurring process in the low penumbra atmosphere. 
A significant amount of magnetic energy may be dissipated through reconnection in the highly abundant PEBs and as such PEBs may play an important role in sunspot decay. 
Outward moving magnetic elements that leave the penumbra and migrate through the sunspot moat, commonly referred to as moving magnetic features (MMF), carry net flux away from the sunspot and are traditionally regarded as main actors in sunspot decay
\citep[see, e.g.][]{2003A&ARv..11..153S}. 
The ubiquity of PEBs we find here may implicate that some fraction of magnetic energy is already dissipated and lost from the sunspot before MMFs cross the sunspot boundary. 
Moreover, high density of EBs in the immediate vicinity of the sunspot suggest that significant fraction of magnetic field in the moat flow regions might also dissipate through magnetic field reconnection occurring in the photosphere. 

What impact do PEBs have on the upper atmosphere? There exist several transient phenomena in sunspots that may be related to energy release in PEBs.
Penumbral micro-jets (PMJ) are short-lived elongated brightenings that can be observed in the core of \ion{Ca}{ii} lines 
\citep{2007Sci...318.1594K, 
2013ApJ...779..143R}. 
%
Magnetic reconnection has been suggested as their driver but the idea that they carry high-speed plasma flows as their name suggests has been contested 
\citep{2019ApJ...870...88E, 
2019A&A...626A..62R}. 
They can be observed in transition region diagnostics
\citep{2015ApJ...811L..33V, 
2020A&A...638A..63D} 
and
\citet{2016ApJ...816...92T} 
reported the existence of large PMJs originating from the outer penumbra in the regions with abundant mixed polarities. These large PMJs leave signatures in some of the transition region/coronal channels of the AIA instrument of NASA's Solar Dynamics Observatory
\citet{2017A&A...602A..80D} 
found that there exist on average 21 PMJs per time step in a time series of \ion{Ca}{ii}~8542~\AA\ observations. This is significantly fewer than the number of PEBs that we detect. Furthermore, clear PMJ detections are mostly found in the inner penumbra where we find fewer PEBs as compared to the outer penumbra.
Recently, 
\citet{2019ApJ...876...47B} 
and \citet{2020A&A...638A..63D} 
connected \ion{Ca}{ii}~8542~\AA\ PMJs with dark fibrilar 
structures close to the line core in \Halpha.
The connection between PEBs and PMJs warrants further study and requires simultaneous observation of multiple spectral lines and extreme high temporal evolution to resolve the onset of PMJs
\citep{2019A&A...626A..62R}. 
Possibly there is also a connection with transition region bright dots observed above sunspots with IRIS 
\citep{2014ApJ...790L..29T, 
2017ApJ...835L..19S} 
and Hi-C \citep{2016ApJ...822...35A}. 
Furthermore, magnetic reconnection in the deep atmosphere as marked by PEBs may play a role in the heating of bright coronal loops that are rooted in penumbrae 
\citep[see, e.g.,][]{2017ApJ...843L..20T} 

Finally, we conclude that EBs in the penumbra of sunspots are an excellent target for new telescopes such as the 4-m DKIST
\citep{2020SoPh..295..172R} 
and the planned
EST \citep{2019arXiv191208650S} 
since PEBs offer opportunities to study magnetic reconnection in kG magnetic field environments at the smallest resolvable scales in astrophysical plasmas.

\begin{acknowledgements}
The Swedish 1-m Solar Telescope is operated on the island of La Palma
by the Institute for Solar Physics of Stockholm University in the
Spanish Observatorio del Roque de los Muchachos of the Instituto de
Astrof{\'\i}sica de Canarias.
The Institute for Solar Physics is supported by a grant for research infrastructures of national importance from the Swedish Research Council (registration number 2017-00625).
This research is supported by the Research Council of Norway, project number 250810, and through its Centres of Excellence scheme, project number 262622.
VMJH is supported by the European Research Council (ERC) under the European Union’s Horizon 2020 research and innovation programme (SolarALMA, grant agreement No. 682462).
%
We thank Shahin Jafarzadeh, Ainar Drews, Tiago Pereira and Ada Ortiz for their help with the observations. 
We made much use of NASA's Astrophysics Data System Bibliographic Services.
\end{acknowledgements}


\end{document}